\documentclass[twocolumn]{revtex4-1}

\usepackage{subcaption}
\usepackage{color}
\usepackage{amsmath}
\usepackage{pgfplots}
\usepackage{tikz}
\usepackage{physics}
\usepackage{float}

\usetikzlibrary{external}
\usetikzlibrary{arrows}
\pgfplotsset{compat=newest}
\begin{document}

\title{Transition temperature scaling in weakly coupled two-dimensional Ising models}

\author{Jordan C. Moodie}
\email{jxm276@student.bham.ac.uk}
\author{Manjinder Kainth}
\author{Matthew R. Robson}
\author{M.W. Long}
\affiliation{
School of Physics and Astronomy, University of Birmingham, Edgbaston, Birmingham, B15 2TT,
United Kingdom.
}
\date{\today}
\begin{abstract}
We investigate the proposal that for weakly coupled two-dimensional magnets the transition temperature scales with a critical exponent which is equivalent to that of the susceptibility in the underlying two-dimensional model, \( \gamma \).
Employing the exact diagonalization of transfer matrices we can determine the critical temperature for Ising models accurately and then fit to approximate this critical exponent.
We find an additional logarithm is required to predict the transition temperature, stemming from the fact that the heat capacity exponent \( \alpha \) tends to zero for this Ising model, complicating the elementary prediction.
We believe that the excitations of the transfer matrix correspond to thermalized topological excitations of the model and find that even the simplest model exhibits significant changes of behavior for the most relevant of these excitations as the temperature is varied.
\end{abstract}

\maketitle
\section{Introduction}
Statistical mechanics is mathematically dominated by phase transitions which are in turn dominated by power laws in the critical region; scaling theory and critical exponents \cite{Fisher1967b, Fisher1974, Hilfer1992, Li2001}.
Universality then suggests that there are only a few styles of phase transitions, which are characterized by critical exponents.
Experimentally these exponents should be measured and then the universality class to which the phase transition belongs can be determined.
Theoretically, the simplest model from each class may be examined and the critical exponents found in order to compare with experiment. 
There are a few exactly solvable models for which the critical exponents are known mathematically exactly \cite{Onsager1944,Kaufman1949,Lieb1964,Wu1982}, but usually numerical procedures are required to approximately determine the exponents \cite{Bartelt1986,Xavier1998,Ghaemi2002}.
This can prove a surprisingly difficult task.

The physical interest for our investigation comes from two-dimensional bilayer magnetic systems.
The transition temperature depends on the coupling between layers and particular scaling arguments are thought to apply when it may be considered weak.
Each layer may be thought of as an effective field acting on the other and hence it is believed that the transition temperature scales with the magnetic susceptibility critical exponent \( \gamma \) of the underlying two-dimensional model \cite{Abe1970,Suzuki1971}.
This surprising relationship has previously been tested with apparent success, in particular by Lipowski for the Ising model \cite{Lipowski1993,Lipowski1997}.
We have been developing an accurate method for determining transition temperatures and so decided it should be employed to reproduce the numerical derivation of \( \gamma \).

Statistical mechanics is a subject where making accurate predictions, either analytically or numerically, is surprisingly challenging.
The weakly coupled pair of square lattice Ising planes has a rich history \cite{Oitmaa1975,Capehart1976,Binder1974,Angelini1995,Monroe2004,Mirza2003} and the particular case where the bonds within each lattice are the same has had several predictions for the transition temperature; Fisher \cite{Weng1967,Fisher1967a}, the M\"uller-Hartman--Zittartz method \cite{MullerHartmann1977, Burkhardt1978}, and Suzuki \cite{Lipowski1993}.
We make an even more accurate prediction for this transition temperature.

In section \ref{sec:technique} we will introduce the numerical technique to find transition temperatures, based on transfer matrices.
This involves solving finite systems to machine precision, typically 15 decimal digits, and extrapolating the results.
Next in section \ref{sec:results} we shall discuss the results generated by this technique for a particular bilayer model.
We will find evidence of a logarithmic correction in the scaling relation and present reasoning for its existence.
Section \ref{sec:comparison} will see us apply our technique to a model more widely studied and hopefully answer the question as to why previous studies did not detect this logarithmic correction.
Finally in section \ref{sec:modeling} we will present interesting physical interpretations of the numerical data we obtain, in the form of topological excitations, which may well be the most important result of this work.
\section{The technique}\label{sec:technique}
In this section we will describe the key numerical technique of this work.
We construct transfer matrices which solve a series of 1D Ising models, each of which contain a system size parameter.
We then use exact diagonalization to find the largest two eigenvalues of these matrices and extrapolate our results to infinite system size, which corresponds to a 2D Ising model.
The transition temperature of the 2D model can then be found approximately.
This, it will turn out, is surprisingly accurate - everything bar the polynomial extrapolation is exact to machine precision and we obtain the transition temperature to many decimal places.

\begin{figure}[t]
  \includegraphics[width=8cm]{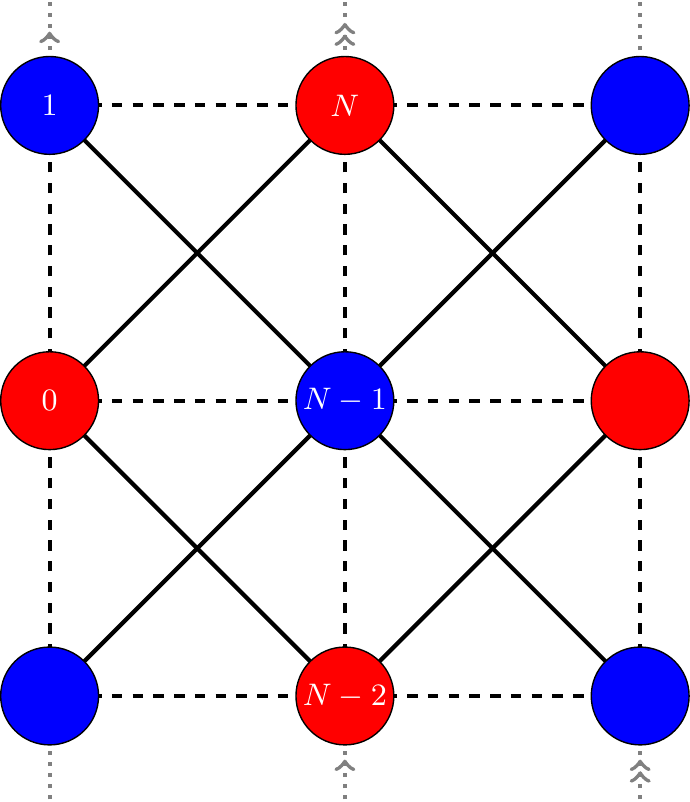}
  \caption{
    Depiction of helical boundary conditions as described in the text.
    In the thermodynamic limit this leads to the square lattice with nearest neighbor (dashed lines) and second-nearest neighbor (solid lines) interactions, as described by equation \eqref{eq:squareLattice}.
    This may be thought of as a two-layered system, coupled by the dashed bonds, though otherwise the red and blue sites are identical.
  } \label{fig:lattice}
\end{figure}

Consider the one-dimensional Hamiltonian
\begin{equation} \label{eq:spinHamiltonian}
  H = -\sum_{n=1}^N a_{n} \sum_j \sigma_j \sigma_{j+n} \,,
\end{equation}
where we have chosen \( \sigma_j \in \{ 1, -1 \} \) to be Ising spins.
Through appropriate choice of the coupling constants \( a_n \), upon taking the limit \( N \to \infty \) this may be used to probe two-dimensional geometries.
Setting \( a_1 = a_{N-1} \equiv J_1 \) along with \( a_{N-2} = a_{N} \equiv J_2 \), with all other matrix elements vanishing, leads to the square lattice with nearest and second-nearest neighbor interactions
\begin{equation} \label{eq:squareLattice}
  H = -J_1 \sum_{\langle jj'\rangle_1} \sigma_j \sigma_{j'} - J_2\sum_{\langle jj'\rangle_2} \sigma_j \sigma_{j'} \,.
\end{equation}
A depiction of the model prior to the thermodynamic limit being taken is provided in Fig. \ref{fig:lattice}.

This application of helical boundary conditions introduces an infinitesimal spiral into our system which has no effect on bulk quantities in the thermodynamic limit.
In practice this means that we may interrogate the two-dimensional system by extrapolating from relatively modest system sizes.
The advantage to these boundary conditions over the more common cylindrical geometry is the sparseness of the Hamiltonian matrix; there are only as many elements in a given row as spin degrees of freedom, 2 for the Ising model.
While the matrix itself is larger as there are fewer symmetries to extract, it is computationally faster to deal with sparse rather than dense matrices \cite{Nightingale1976,Nightingale1977,Nightingale1980}.
One symmetry which does remain, however, is the global spin symmetry.
For example, consider the groundstate for the Ising model, where all spins are aligned.
Whether their orientation is labeled up or down is irrelevant.
In practice this irrelevance allows us to halve the state space, and thus the size of any matrix, and remove the double groundstate.
On a technical level this is done by using the variables \( \tau_j = \sigma_j \sigma_{j+1} \), which describe the relative orientation of each spin.

\begin{figure*}[t!]
  \begin{subfigure}{.45\textwidth}
    \includegraphics[width=8cm, height=8cm]{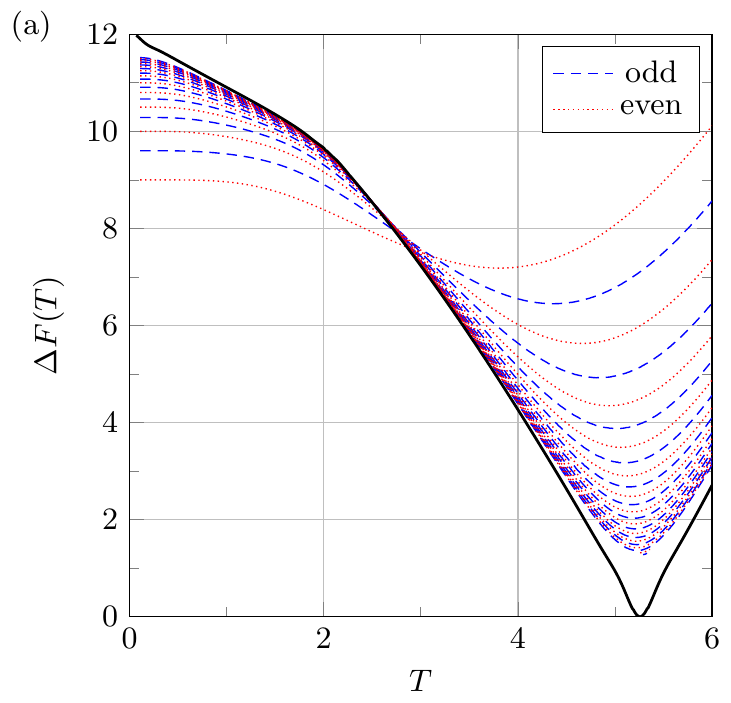}
  \end{subfigure}
  \begin{subfigure}{.45\textwidth}
    \includegraphics[width=8cm, height=8cm]{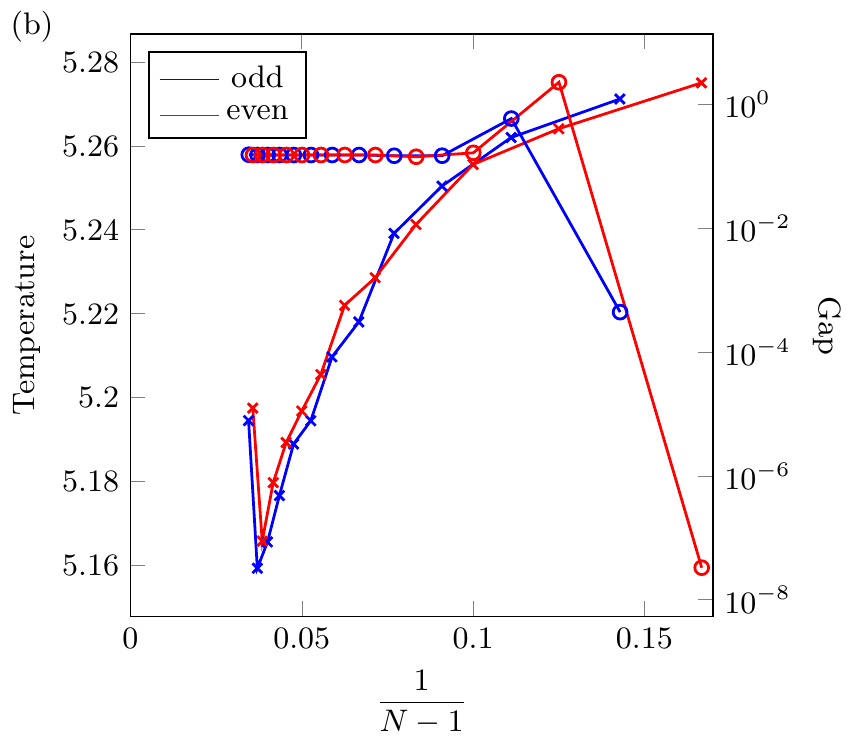}
  \end{subfigure}
  \caption{ 
    Parameters for both graphs are \( \lambda = 1 \) for \( N = 4\text{-}29 \).
    a) The free energy difference, \( \Delta F(T) \).
    The odd curves are blue and dashed, while the even curves are red and dotted.
    The solid line is the extrapolation to infinity from the odd curves.
    For the larger systems only data near the transition temperature has been calculated.
    b) Circles denote the temperature at which the free-energy difference is minimum for extrapolations using increasing larger systems, and crosses denote the gap at this point.
    The increase in the gap for the final two points is explained in the text.
  } \label{fig:J1J2-1.00-combined}
\end{figure*}

From this Hamiltonian we construct a transfer matrix,
\begin{equation}
  \hat{T} = \hat{t}^{N-1} \equiv e^{-\beta \hat{F}} \,,
\end{equation}
where explicitly
\begin{multline}
  \bra{\sigma_0,\, \sigma_1,\, \dots,\, \sigma_N} \hat{t} \ket{\sigma_0',\, \sigma_1',\, \dots,\, \sigma_N'} \\
  =e^{- \beta \lambda (\sigma_0\sigma_1 + \sigma_0\sigma_{N-1} )
  - \beta ( \sigma_0\sigma_{N-2} + \sigma_0\sigma_N ) } \prod_{j=0}^{N-1} \delta_{\sigma_{j+1},\, \sigma_j'}\,,
\end{multline}
and the ratio \(\lambda \equiv J_1 / J_2 \) sets the relative strength of the two bonds.
We will be primarily interested in the scaling behavior as \( \lambda \to 0 \) where the lattice depicted in Fig. \ref{fig:lattice} may be thought of as two weakly-coupled square lattices.
The submatrix \( \hat{t} \) acts to ratchet around the spiral to the next spin site (from 0 to 1, for example) while the transfer matrix \( \hat{T} \) transfers to the site right of our start position (from 0 to \(N-1\)).
The free-energy operator \( \hat{F} \) is interpreted as a quantum mechanical Hamiltonian which in the thermodynamic limit is one-dimensional.
Standard ideas concerning spectra of Hamiltonians, in particular energy gaps, will then apply to this operator with energy replaced by free-energy.
We may define the free-energy gap in terms of eigenvalues of the submatrix \( \hat{t} \),
\begin{equation}\label{eq:DeltaFm}
  \Delta F_m = \frac{N-1}{\beta} \ln \frac{t_0}{t_m} \,,
\end{equation}
where \( t_0 \) is the largest eigenvalue and hence becomes the partition function, while \( m \) labels the eigenvalues in the symmetric subspace in order.

We calculate the largest two eigenvalues of this submatrix using the power method, for a variety of system sizes.
From these we may obtain \( \Delta F_1(N,\,T) \).
We then employ polynomial extrapolation for these finite systems at a fixed temperature to effectively obtain \( \Delta F_1(\infty,\, T) \equiv \Delta F(T) \).
This is directly analogous to the finite-size scaling of exact diagonalization results common in the quantum mechanics literature.
However, as we will see, it performs much better in this thermodynamic context.
As mentioned earlier, the finite data is exact to machine precision and so the only errors come from the polynomial extrapolation.

We interpret the function \( \Delta F(T) \) as describing how the free-energetic cost of a topological excitation changes with temperature, while the other \( \Delta F_m(T) \) describe topological excitations of higher cost.
An example for the \( J_1 \)-\( J_2 \) model previously described is given in Fig. \ref{fig:J1J2-1.00-combined}a.
We shall describe the exact nature of this excitation in greater detail in later sections. 
For now it is sufficient to understand that when the cost of these excitations becomes zero, the system undergoes a phase transition.
This can be easily understood for an Ising model where the transition is defined by the change from the low-temperature ordered phase, where there is a single divergent cluster of one spin orientation, to a high-temperature disordered phase.
The topological excitations in this case are domain walls; as the transition temperature is approached their energetic cost is increasingly compensated for by entropic gain, until both exactly balance and macroscopic domain walls are permitted.

Using this idea, the transition temperature of a model is defined as the temperature at which the free-energetic cost of a topological excitation is zero
\begin{equation}\label{eq:TcDefinition}
  \Delta F (T_c) \equiv 0 \,.
\end{equation}
We then use this definition to find the transition temperature of a variety of models.

In practice we may only approximate \( \Delta F (T) \) by extrapolating a moderate number of finite systems.
Figure \ref{fig:J1J2-1.00-combined}a demonstrates this idea for the case \( \lambda = 1 \), when the intra- and inter-layer bonds are equal.
The nature of this approximation means that any points of non-analyticity, for example at the transition temperature, will instead appear analytic.
This is seen in the plot where near the transition temperature the extrapolation is quadratic.
Increasing system size reduces the region in which it is quadratic, until it becomes indistinguishable from a cusp on a fixed finite temperature scale.
As it is just an approximation, the minimum of this quadratic will not be zero as demanded by equation \eqref{eq:TcDefinition}.
However, the more accurate the approximation the closer to zero the minimum will be.
It is this gap between the minimum and zero that we will use as a measure of the accuracy of our technique.

Figure \ref{fig:J1J2-1.00-combined}b demonstrates this idea, where we see the gap shrink as more systems are used to extrapolate, with the exception of the extrapolation using the largest systems.
This anomaly is due to reaching a numerical limit.
We would have to perform calculations using data types which have more than the standard 15 decimal digits of precision if we wished to reduce the gap beyond \( 10^{-8} \).
In general we see the transition temperature oscillate within some exponentially converging envelope as the number of systems is increased.
As the estimate of the transition temperature converges, the gap rapidly decreases.
Note that the accuracy of the transition temperature is approximately equal to the gap.

\begin{figure}[h]
  \includegraphics[width=8cm, height=8cm]{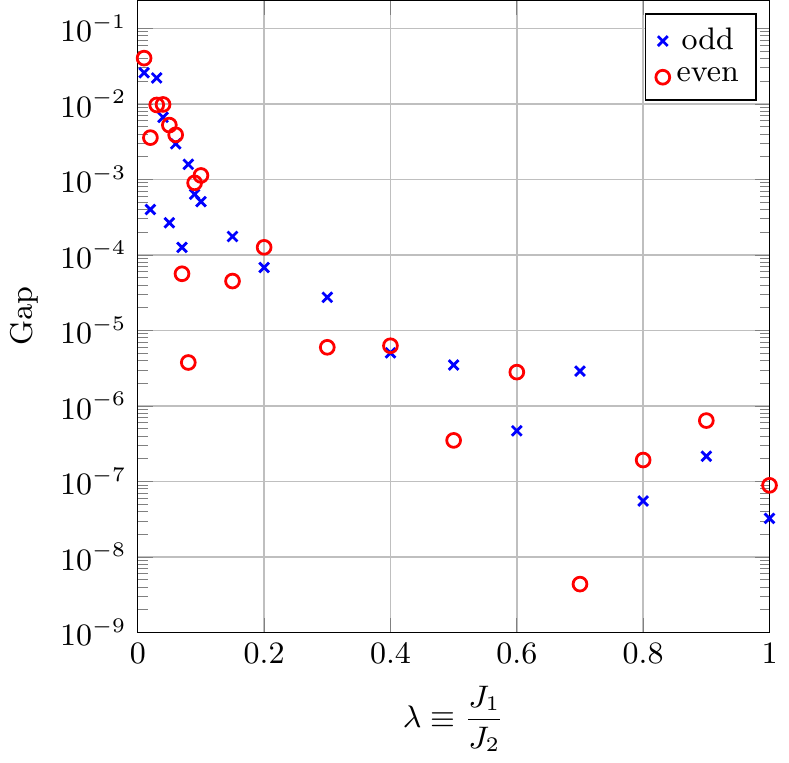}
  \caption{ 
    The gap for a variety of coupling strengths.
    Blue crosses denote extrapolations using odd sized systems while red circles use even sized systems. 
    In each case we use the extrapolation which has the lowest gap, which is generally the largest systems.
  } \label{fig:J1J2errors}
\end{figure}

We are interested in probing the scaling behavior as \( \lambda \to 0 \), when the square lattice with nearest and second-nearest neighbor interactions may be thought of as two weakly-coupled square lattices.
In Fig. \ref{fig:J1J2errors} we show how the gap depends on the bond ratio \( \lambda \).
Note that at each point we use the extrapolation which produces the smallest gap.
This typically contains the largest system (\( N=28,\,29 \)) but for larger \( \lambda \), like in Fig. \ref{fig:J1J2-1.00-combined}b, it instead tends to only contain data up to \( N = 26,\,27 \).
There are two things to note in this picture.
First we find that convergence worsens dramatically in the low \( \lambda \) limit.
Second, we see that in this limit odd systems tend to give better results.
While they do benefit from containing the largest calculation we have performed (\( N = 29 \), as opposed to 28 for even systems), odd data generically converges faster than even data.
The reasons for each of these phenomena are subtle and relate to the aforementioned topological excitations.
The first involves the excitations changing behavior near the transition temperature for low \( \lambda \), while the second is due to even-specific variants.
As such, we shall leave further explanation of this observation to section \ref{sec:modeling}.

\section{Results \& Analysis}\label{sec:results}

We now turn to using the data generated from the technique outlined in the previous section to investigate a very interesting scaling relation.
For a bilayer system, the presence of a second Ising layer increases the tendency of spins to order within each layer, raising the transition temperature past that of the uncoupled case.
In essence, each layer acts as an effective field on the other.
For this reason it has been beautifully argued \cite{Abe1970,Suzuki1971} that the change in the transition temperature scales with the critical exponent \( \gamma \), associated with the magnetic susceptibility.
This is at first rather surprising, but subsequent numerical investigations seemed to confirm it \cite{Lipowski1993,Lipowski1997}.

More precisely, let us assume the singular part of the free energy for some model with a transition temperature \( T_c(\lambda) \) is given by
\begin{equation}\label{eq:generalScaling}
  F_\lambda^\text{sing} \sim t_\lambda^{2-\alpha} \qquad \text{as } t_\lambda \to 0 \,,
\end{equation}
where \( t_\lambda = [T-T_c(\lambda)]/T_c(\lambda) \) is the reduced temperature and \( \alpha \) a critical exponent.
For two weakly coupled layers the singular part of the free energy was previously calculated in terms of the uncoupled case.
To leading order this is given by \cite{Abe1970}
\begin{equation}\label{eq:Abe}
  F^\text{sing} \sim F_0^\text{sing} - a t_0^{2-\alpha}\left( \frac{\lambda}{t_0^\gamma} \right)^2 \,,
\end{equation}
where the first order term vanishes due to spin symmetry.
Equations \eqref{eq:generalScaling} and \eqref{eq:Abe} may then be expected to balance at the transition temperature \( T_c(\lambda) \).
Exactly at this point \( F_\lambda \) is zero and so 
\begin{equation}
  \tau^{2-\alpha} = a \tau^{2-\alpha}\left( \frac{\lambda}{\tau^\gamma} \right)^2 \,,
\end{equation}
where \( \tau ( \lambda) \equiv [T_c(\lambda) - T_c(0)] / T_c(0) \).
This \( \tau (\lambda) \) is the reduced temperature of the monolayer model, \( t_0 \), exactly at the transition temperature of the bilayer model, \( T_c(\lambda) \).
This equation immediately gives
\begin{equation}\label{eq:existingFormula}
  \lambda \sim \tau^\gamma \,.
\end{equation}

\begin{figure}
  \includegraphics[width=8cm, height=8cm]{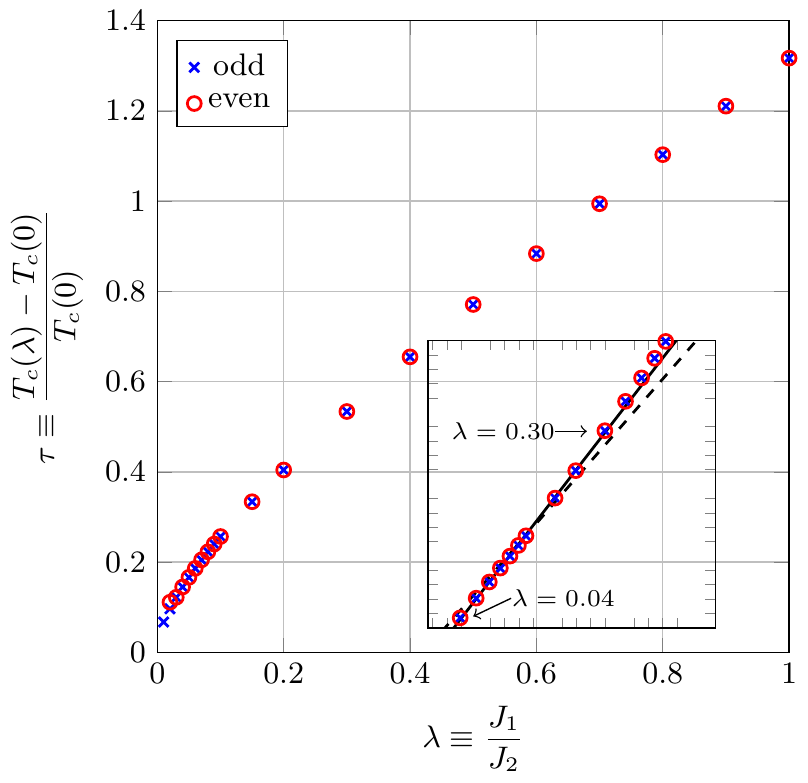}
  \caption{ 
    Dependence of the transition temperature on coupling strength.
    The inset shows a log-log plot of data in the range \(  0.04 \leq \lambda \leq 0.30 \), and two lines are then fit to the odd systems.
    The solid line is a linear fit, from whose slope it can be determined that \( \gamma \approx 1.55 \).
    The dashed line is a fit assuming the slope must be given from \( \gamma = 7/4 \).
  } \label{fig:existingFormula}
\end{figure}

We turn now to analyzing the data to test this hypothesis.
Figure \ref{fig:existingFormula} shows how the transition temperature of the \( J_1 \)-\( J_2 \) system changes with coupling strength. 
In particular, the inset displays the logarithm of this data.
It should be possible then to calculate \( \gamma \) from the slope of this picture and we attempt to do so.
Unfortunately, a least-squares fit finds \( \gamma \approx 1.5 \) which does not agree with the known value of \( \gamma = 7 / 4 \) for the square-lattice Ising model.
Indeed, fitting a line using this known value does not lead to an acceptable fit.
Additionally, the logarithm of the data displays clear curvature which implies that a simple power law is not correct.

\begin{figure}
  \includegraphics[width=8cm, height=8cm]{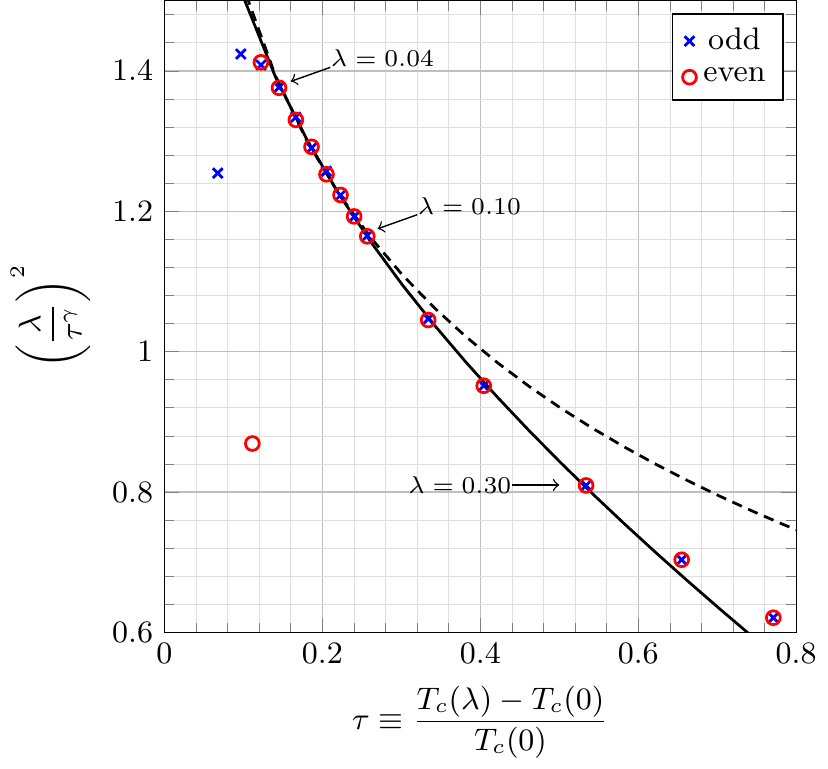}
  \caption{ 
    Fits to the proposed new formula \eqref{eq:newFormula}, assuming \( \gamma = 7/4 \).
    The dashed line fits to odd systems in the range \(  0.04 \leq \lambda \leq 0.10 \) while the solid line includes a linear correction (that is, including an additional term \( b \tau \))  and fits to odd systems in the range \(  0.04 \leq \lambda \leq 0.30 \).
  } \label{fig:newFormula}
\end{figure}

This failure of fitting is caused by a subtlety of the Ising model.
In this model \( \alpha \to 0 \) and so equation \eqref{eq:generalScaling} should be replaced with
\begin{equation}
  F_\lambda^\text{sing} \sim t_\lambda^2 (\log t_\lambda - a ) \qquad \text{as } t_\lambda \to 0 \,,
\end{equation}
and consequently we find
\begin{equation}\label{eq:newFormula}
  \left( \frac{\lambda}{\tau^\gamma} \right)^2 \sim a - \log \tau \,.
\end{equation}
We attempt to least-squares fit this formula in Fig. \ref{fig:newFormula}.
The fit is remarkably good for low \( \lambda \) and a linear correction allows it to be extended further from the critical region.
Including the logarithm improves the fit by orders of magnitude compared to any polynomial attempts.
It should be noted that if equation \eqref{eq:existingFormula} were correct then Fig. \ref{fig:newFormula} would be flat as the function being plotted would be a constant.
This then provides a sensitive measure of the accuracy of the data.
The revised formula also immediately explains why the previous fit found an incorrect exponent; when fitting a dominant power-law, a logarithmic term acts to corrupt its exponent.

This evidence leads us to conclude that there is indeed a remarkable scaling relationship between the transition temperatures of bilayer systems and the magnetic susceptibility of their monolayer counterparts, but that for the Ising model it is complicated by a logarithm.

\section{Comparison with another model}\label{sec:comparison}

\begin{figure}
  \includegraphics[width=8cm]{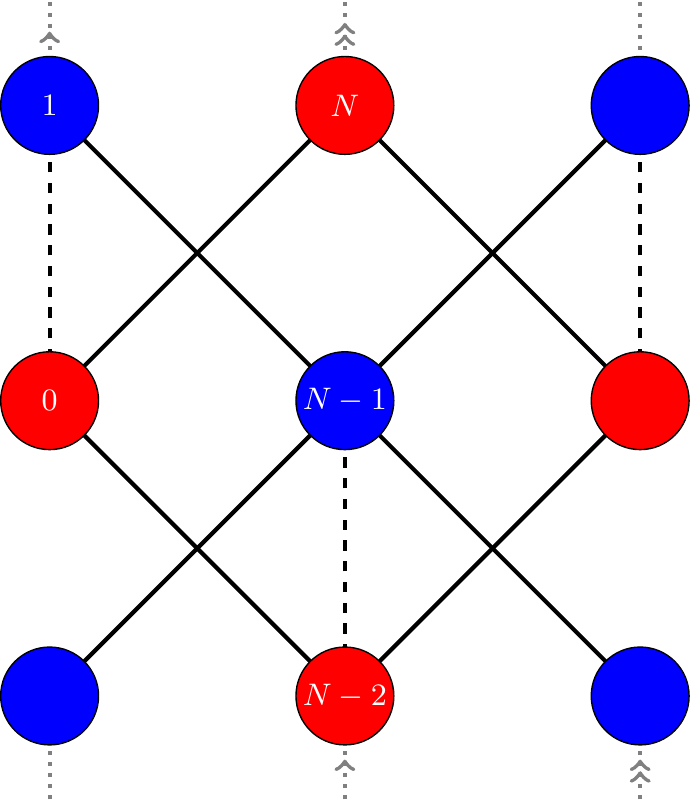}
  \caption{
    A depiction of what becomes the cubic two-layer model in the thermodynamic limit.
    The solid lines denote in-layer bonds while the dashed lines denote between-layer bonds.
  } \label{fig:cubicLattice}
\end{figure}

Thus far we have dealt with the \( J_1 \)-\( J_2 \) model, the main benefit of which is that all sites are equivalent so it is simple to apply our technique.
However, the model most often seen in related literature is one where the second lattice sits directly above the first, creating a more cubic structure \cite{Lipowski1993,Lipowski1997,Oitmaa1975}.
This is depicted in Fig. \ref{fig:cubicLattice}.
The scaling laws described in the previous section should be universal, provided the models remain local, and so we thought that it would be worthwhile to test our technique on the more standard geometry.

Unlike the \( J_1 \)-\( J_2 \) model, this cubic structure necessitates two atoms per unit cell.
The transfer matrix is then more complicated; we must construct two one-dimensional transfer matrices which are applied in succession to carry us between equivalent sites.
This means that only even sized systems exist.

\begin{figure}[h!]
  \includegraphics[width=8cm, height=8cm]{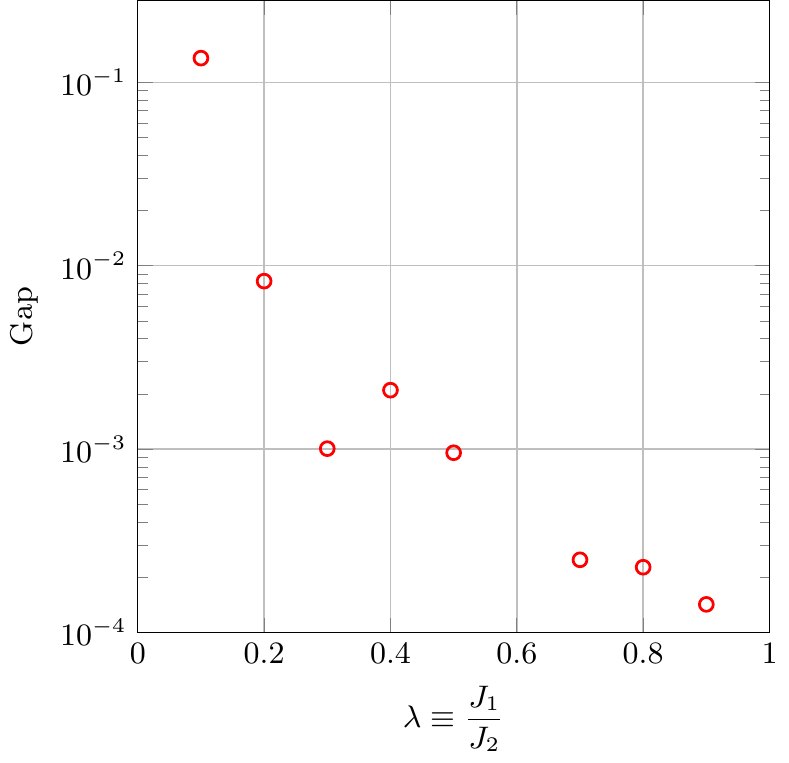}
  \caption{ 
    The gap of the cubic two-layer model for a variety of coupling strengths.
    The data comes from extrapolations for even sized systems with \( N = 4\text{-}28 \).
  } \label{fig:cubicErrors}
\end{figure}

\begin{figure}[h!]
  \includegraphics[width=8cm, height=8cm]{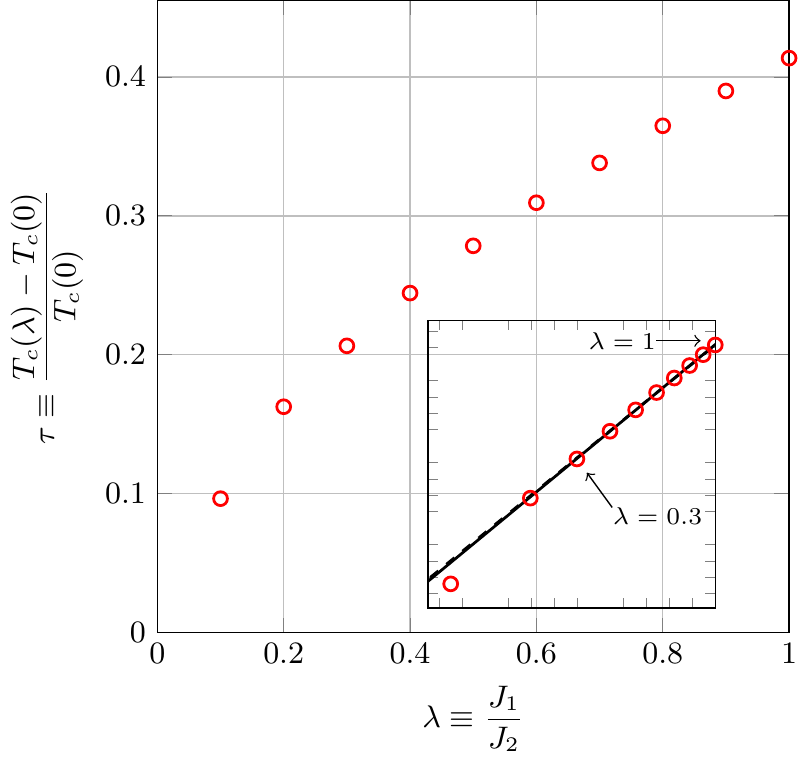}
  \caption{ 
    Dependence of the transition temperature on coupling strength for the cubic two-layer model.
    The inset shows a log-log plot of the data and two lines are fit to data in the range \( 0.3 \leq \lambda \leq 1 \).
    The solid line is a simple linear fit from whose slope it can be determined that \( \gamma \approx 1.73 \).
    The dashed line is a fit assuming the slope must be given from \( \gamma = 7/4 \).
  } \label{fig:cubicExistingFormula}
\end{figure}

\begin{figure}[h!]
  \includegraphics[width=8cm, height=8cm]{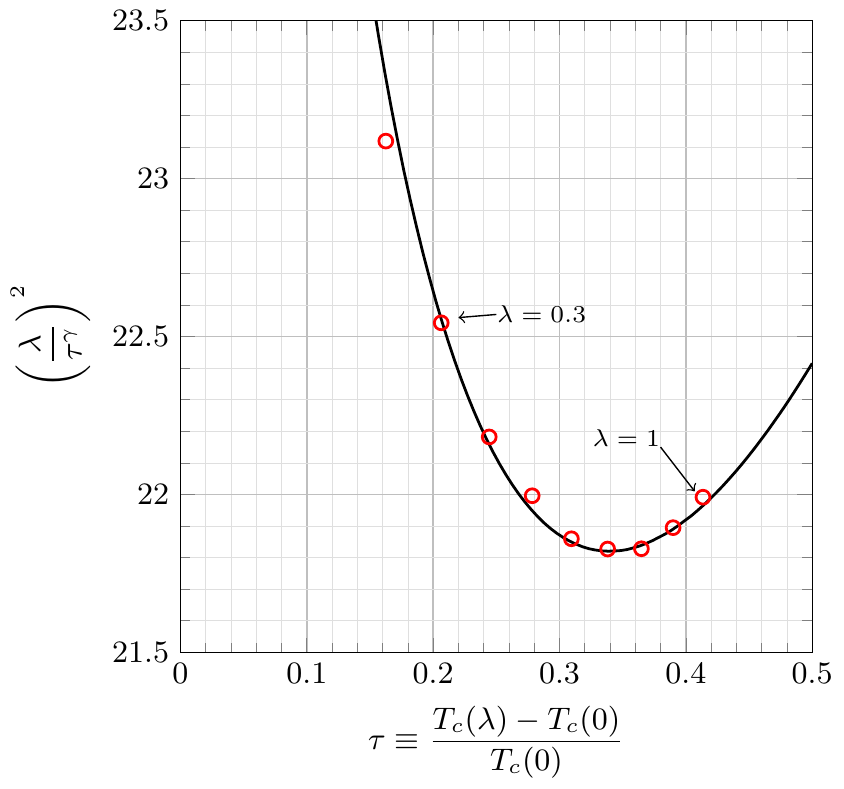}
  \caption{ 
    A fit to the proposed new formula \eqref{eq:newFormula}, with a linear correction and assuming \( \gamma = 7/4 \).
    We fit to systems in the range \( 0.3 \leq \lambda \leq 1 \).
  } \label{fig:cubicNewFormula}
\end{figure}

\begin{figure}[h!]
  \includegraphics[width=8cm, height=8cm]{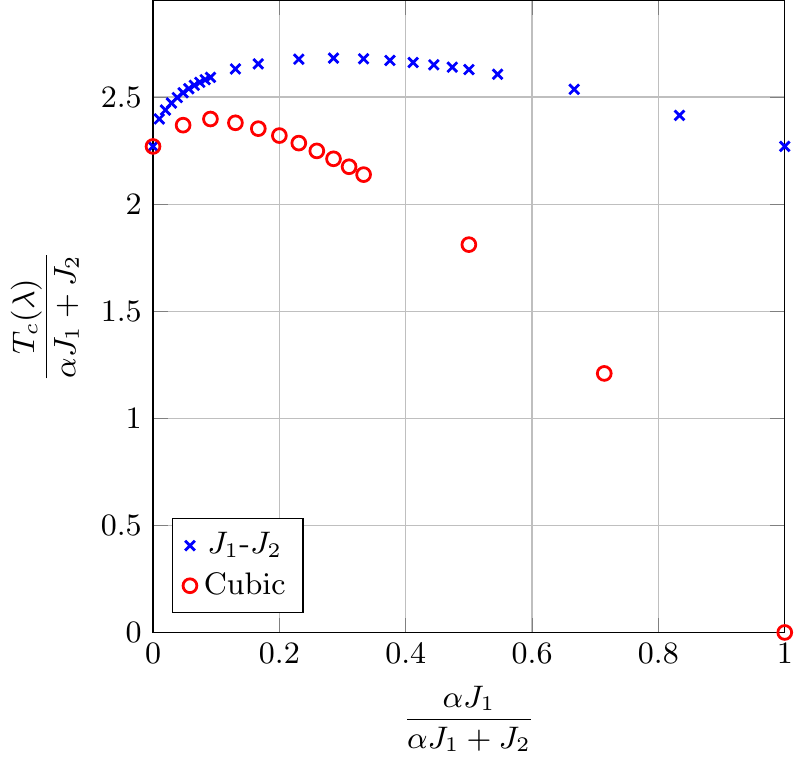}
  \caption{ 
    How the scaled transition temperature depends on scaled bond-strength for both the \( J_1 \)-\( J_2 \) and cubic  two-layer models.
    The parameter \( \alpha \) is \( 1 \) for former as there is an equal number of both \( J_1 \) and \( J_2 \) bonds, while for the latter it is is \( 1/2\) as there are twice as many \( J_2 \) bonds as \( J_1 \) bonds. 
    The points at \( J_1 = 0 \) and \( J_2 = 0 \) are exact, as described in the text.
  } \label{fig:twoModel}
\end{figure}

\clearpage

As shown in Fig. \ref{fig:cubicErrors}, the errors intrinsic to this model are much larger and begin at much higher coupling strength \( \lambda \) that the previous Fig. \ref{fig:J1J2errors}.
Any fitting must thus be performed further away from the weak-coupling limit than may be preferred.

In Fig. \ref{fig:cubicExistingFormula} we plot how the transition temperature of this new cubic model changes with coupling strength.
The inset shows the logarithm of the data.
The linear fit to this logarithm, testing equation \eqref{eq:existingFormula}, gives a reasonable approximation to the critical exponent \( \gamma \), namely \( \gamma \approx 1.73 \).
The line constructed using the known \( \gamma = 7/4 \) is not noticeably different in the region being fitted, though the data does display observable curvature.

Figure \ref{fig:cubicNewFormula} instead tests the proposed scaling relation \eqref{eq:newFormula} with a linear correction.
Again it must be noted that this plot would be expected to be flat if equation \eqref{eq:existingFormula} were correct.
If a technique only had access to data near the minimum of this curve then one may be led to believe that it is indeed flat and hence that the previous formula held.
This then may explain its apparent corroboration by previous numerical investigations.
In our case the absence of such a flat region for the \( J_1 \)-\( J_2 \) model precluded such a conclusion.
The fit to the new formula of course performs admirably, except the point for \( \lambda = 0.1 \) which has not converged.
It should be noted that, compared to the previous section, the linear correction here is sizable and crucial to the fit.

These two facts, non-convergence of the transition temperature for comparably large \( \lambda\) along with a sizable linear correction, may suggest that the critical region for this cubic model is much smaller than that of the \( J_1 \)-\(J_2\) model.
As such the fit obtained in Fig. \ref{fig:cubicExistingFormula} would be coincidental and not expected.
To examine this claim we scale the transition temperatures obtained for the two models with the total bond-strength and plot against scaled bond-strength in Fig. \ref{fig:twoModel}.
Note that the points at which either the between-layer bond \( J_1 = 0 \) or the in-layer bond \( J_2 = 0 \) are exact for both models.
For the \( J_1 \)-\( J_2 \) model both of these cases result in two disconnected square lattices, which can be exactly solved.
This is also the case for the cubic model when \( J_1 = 0 \), but for \( J_2 = 0 \) the system becomes a collection of disconnected dimers which have no transition.
We thus see that the region in which the graph rises is much smaller for the cubic model, perhaps implying the critical region is much more constrained.

\section{Modeling the topological excitations}\label{sec:modeling}

While the previous two sections dealt with the numerical evidence underpinning our proposed correction to a scaling relationship, what follows is unrelated to any scaling theory.
Instead here we shall attempt to explain the physical meaning of quantities calculated by the numerical technique in section \ref{sec:technique}.
By doing so we hope that some of the subtleties contained within the numerics will become clear, in particular the difference in accuracy between firstly low and high \( \lambda \) models and secondly extrapolations from even and odd sized systems.
We previously claimed that the curves in Fig. \ref{fig:J1J2-1.00-combined}a represented the free-energy cost of a topological excitation.
In this section we will provide some simple justification for this claim.
While we will discuss the \( J_1 \)-\( J_2 \) model in some detail, similar arguments also apply to the cubic model.

Recall from section \ref{sec:technique} that we may think of a transfer matrix in terms of a quantum mechanical Hamiltonian.
In particular, we may write
\begin{equation}
  \hat{T} \equiv e^{-\beta \hat{F}} \,,
\end{equation}
where for our purposes take \( \hat{T} \) to be the transfer matrix associated to the \( J_1 \)-\( J_2 \) model on a cylinder and \( \hat{F} \) the associated free-energy operator.
As the transfer matrix describes a two-dimensional statistical mechanics problem, the free-energy operator should describe a one-dimensional quantum mechanics problem.
We propose that the particles of the latter correspond to topological excitations in the former.
More concretely, these topological excitations are domain walls which propagate parallel to the axis of the cylinder.
By diagonalizing the transfer matrix we are rigorously thermalizing these domain walls.
As such, fluctuations at non-zero temperature are inherently present and lead to interesting physics.

\begin{figure}
  \includegraphics[width=8cm]{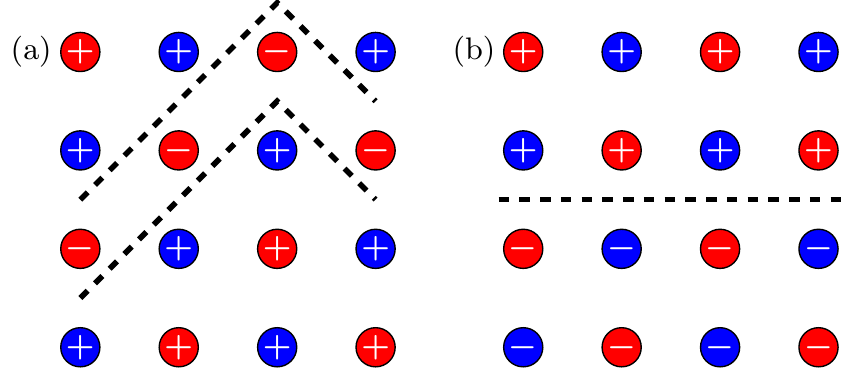}
  \caption{
    Depiction of two styles of topological excitations.
    The red and blue sites indicate different layers, while \( + \) and \( - \) indicates the value of the Ising spin at each site.
    a) A localized excitation, with a line of flipped spins in one layer.
    b) A domain wall running through both layers, separating the system into a region of \( + \) spins and a region of \( - \) spins. For a periodic system these must appear in pairs, though they may be well separated and so be thought of as independent. 
  } \label{fig:excitations}
\end{figure}

Figure \ref{fig:excitations} depicts two natural styles of topological excitations for this model.
The first is a line of flipped spins in just one of the layers.
Every horizontal step breaks two intra- and four inter-layer bonds, resulting in a energy cost of \( 8 J_1 + 4 J_2 \) compared to the groundstate.
The second is a domain wall which cuts through both layers, leaving a region of up spins on one side and down spins on the other.
As this is a periodic system, such domain walls must come in pairs and hence the energy cost of such an excitation is \( 2 ( 2 J_1 + 4 J_2 ) \).
Clearly for low \( \lambda \equiv J_1/J_2 \) the localized excitation is cheaper and so at low temperature would be preferred.
However, this model is in the 2D Ising universality class and hence the phase transition is controlled by excitations of the second type.
As such there must be a crossover between the two as temperature is increased, caused by the aforementioned fluctuations.
In essence we expect thermal fluctuations to cause the localized excitation to spread out over some range, whose average is determined by the temperature, though the edges remain bound.
This costs an energy proportional to \( \lambda \) multiplied by the range and so this cannot continue indefinitely.
Once sufficiently spread out, it is preferable to instead flip some of the spins in the other layer to match those of the excitation.
Such an unbound excitation is then indistinguishable from the independent pair of domain walls, albeit dressed with thermal fluctuations. 
It is this prediction that we will use to test the validity of our assertion.

We can use a simple partition sum argument to model each style of topological excitation.
More sophisticated techniques may be used, for example using the Baker-Campbell-Hausdorff formula, but these are beyond the scope of this paper.
For the localized excitation the free-energy cost may be written as
\begin{equation}\label{eq:localized}
  \Delta F_{\text{local}} = - T \log\left( 2 e^{-\beta (8 J_1 +  4 J_2)} + e^{-\beta ( 4 J_1 + 8 J_2 )} \right) \,.
\end{equation}
The three exponents come from the three directions the excitation can choose to go in a given step: \( 45^\circ \) up, \( 45^\circ \) down, or horizontally across, .
The first two of course have the same energetic cost as we previously discussed, while the final may be thought of as the cheapest fluctuation.
Similarly for the Ising domain wall excitation we may write
\begin{equation}\label{eq:domainwall}
  \Delta F_{\text{Ising}} = - 2T \log \left( e^{-\beta(2 J_1 + 4 J_2)} + 2e^{-\beta (4 J_1 + 4 J_2)} \right) \,.
\end{equation}
Note here the overall factor of two is due to the fact the domain walls must be created in pairs, which we treat as sufficiently separated as to be independent.

\begin{figure}[h!]
  \includegraphics[width=8cm, height=8cm]{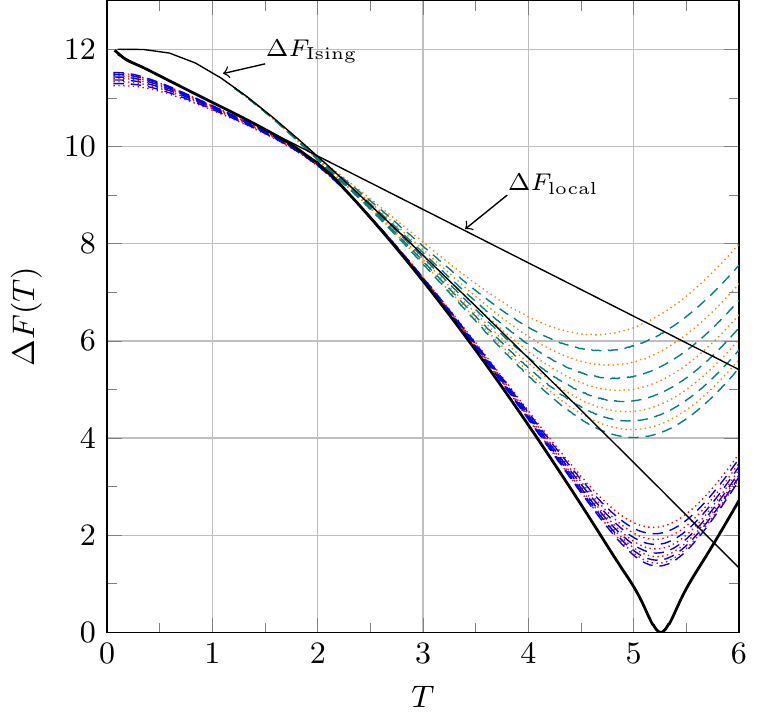}
  \caption{ 
    The free-energy difference for the first and second excitations for the \( J_1 \)-\( J_2 \) model at \( \lambda \equiv J_1 / J_2 = 1 \), where \( N = \) 16-25.
    Odd data is dashed while even data is dotted.
    The solid lines are fits to models of topological excitations described in the text along with an extrapolation to infinity from the odd curves for the first excited state.
  } \label{fig:crossing}
\end{figure}

In Fig. \ref{fig:crossing} we plot the first two excitations, that is \( \Delta F_1 \) and \( \Delta F_2 \) in the language of equation \eqref{eq:DeltaFm}, for the \( J_1 \)-\( J_2 \) system for \( N = \) 16-21 and \( \lambda \equiv J_1/J_2 = 1 \).
We overlay each of the two approximate models given above.
As expected, for very low temperatures the localized excitation model \eqref{eq:localized} fits remarkably well to the first excitation and the Ising domain wall description \eqref{eq:domainwall} to the second.
The two models then cross around \( T \approx 2 \) where indeed we see an avoided crossing in the exact data, indicating a change of behavior.
Beyond this point neither line fits well as both are in fact low-temperature expansions and low-temperature pictures.
To improve the fit more expensive fluctuations would need to be included.
We have tested this modeling for various values of \( \lambda \) and find similar results to that just described.

\begin{figure}[h!]
  \includegraphics[width=8cm, height=8cm]{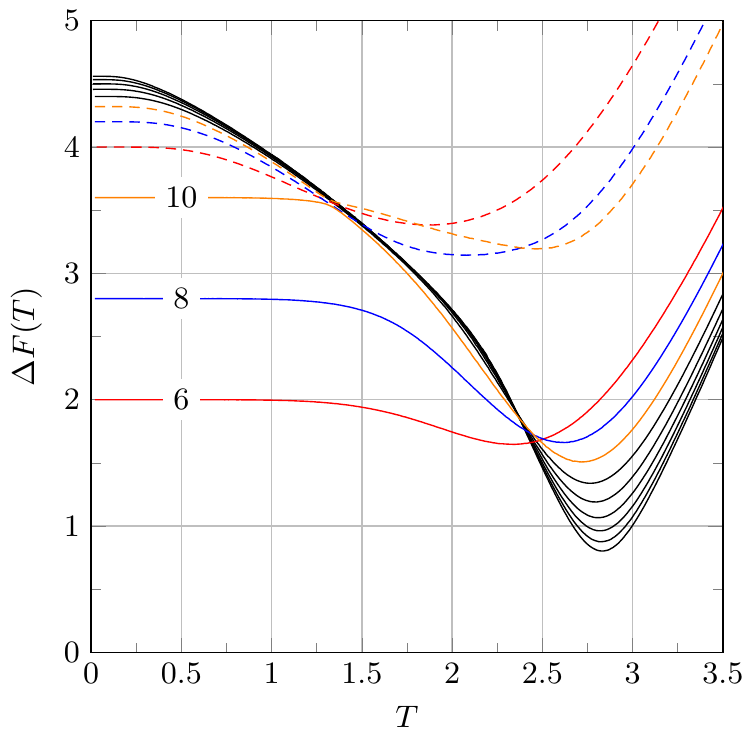}
  \caption{ 
    The free-energy difference for the \( J_1 \)-\( J_2 \) model for \( \lambda \equiv J_1 / J_2 = 0.1 \), for even systems with  \( N = \) 6-22.
    The solid lines are first excitations and the dashed lines second excitations for \( N = 6,\,8,\,10 \).
  } \label{fig:finite}
\end{figure}

Changing the bond ratio \( \lambda \) alters where this crossover takes place.
When \( \lambda \) is small the bonds between layers are weak compared to those within each layer, meaning that the localized excitation can spread out further within one layer before it changes behavior.
This pushes the crossover closer to the transition temperature.
As such it is possible that our finite data still contains information about this local excitation at the temperatures used to extrapolate and estimate the transition temperature of the infinite model.
Indeed, if the length-scale on which the excitation spreads is larger than the system size then this must be the case.
The implication of this is that for low \( \lambda \) larger systems are required to achieve the same level of accuracy obtained for higher values.
This would then explain the convergence problems highlighted in figures \ref{fig:J1J2errors} and \ref{fig:cubicErrors}.

Finally, there is a style of excitation only present in some even sized systems.
These excitations are when one layer is of opposite spin to the other, incurring an energy cost of \( 4 (N-1) J_1 \).
This can be seen in Fig. \ref{fig:finite} for \( \lambda \equiv J_1 / J_2 = 0.1 \), where systems of size 6, 8, and 10 all exhibit this phenomenon.
Initially the free-energy cost of these excitations is constant, as there is no natural way for these to gain entropically, before there is an avoided crossing similar to what was previously described.
After this point the lines behave as usual.
These superfluous, from the point of view of the thermodynamic limit, excitations impact the extrapolation and are the reason odd systems outperformed even systems in accuracy for low \( \lambda \).
Odd systems, of course, have M\"obius boundary conditions, meaning there is only one lattice.

This simple picture of topological excitations is a clear physical interpretation of transfer matrix calculations.
It provides a helpful intuition for numerical results, while concretely explaining otherwise nonobvious facts about convergence.
It is perhaps the most interesting and important aspect of this work and will be the target of future study.

\newpage

\section{Conclusion}
We have presented strong evidence which agrees with the assertion that the transition temperature of weakly-coupled two-dimensional magnets scales with the critical exponent \( \gamma \).
The form of this scaling relation, however, must include a logarithm when dealing with the Ising model.
Detection of this correction was made possible by both the choice of model and the accuracy of the transfer matrix technique.

Transfer matrix calculations have the great advantage of giving the exact answer, to machine precision, for finite systems.
Extrapolation of these systems to the infinite case is then pure, with no noise.
Contrast this with Monte-Carlo methods which can only get approximate answers for slightly larger finite systems.
In addition, we have only taken a very basic approach.
It is possible that if more sophisticated mean-field ideas \cite{Lipowski1993,Lipowski1992} were employed within our technique then even more accurate results may be produced.

Perhaps the most physically interesting aspect to these transfer matrices is the interpretation of the eigenvalues as the free-energy cost of topological excitations.
This interpretation gave an intuitive and surprisingly accurate picture to the low-temperature data.
It also helped explain why certain systems gave less accurate answers than others, giving a practical reason for their study.

To summarize, transfer matrices provide an exceptionally accurate way to perform calculations in statistical mechanics while simultaneously describing incredibly interesting physics. 

\bibliography{bib}{}

\end{document}